\documentstyle[preprint,eqsecnum,aps]{revtex}
\begin{document}
\draft
\title{Background thermal contributions in testing the Unruh effect}
\author{Sandro S.\ Costa and George\ E.A.\ Matsas}
\address{
Instituto de F\'\i sica Te\'orica, Universidade Estadual Paulista\\
R. Pamplona 145, 01405-900 - S\~ao Paulo, S\~ao Paulo, Brazil}
\maketitle


\begin{abstract}
We consider inertial and accelerated Unruh-DeWitt detectors
moving in a
background thermal bath and calculate their excitation rate.  It
is shown that for fast moving detectors such a thermal bath does not
affect substantially the excitation probability.  Our results
are discussed in connection with a possible proposal of testing
the Unruh effect in high energy particle accelerators.
\end{abstract}
\narrowtext
\section{Introduction}
\label{sec:intro}
It is already two decades since Hawking discovered the striking
result that quantum mechanics may induce black holes to
evaporate \cite{H}.  Many different questions related with such an effect
have been clarified since that time, and a number of other ones
are presently under investigation \cite{B}. The so-called Unruh effect
\cite{U,D} has played an outstanding role in emphasizing some of
the exquisite features present in the Hawking effect. According
to the Unruh effect, a detector uniformly accelerated through
the {\em inertial}
vacuum  responds as being in a thermal bath characterized by a
temperature proportional to its proper acceleration. We call
{\em inertial} vacuum the no-particle state as described by
a family of observers following an {\em inertial} timelike
Killing field $\partial_t$ in Minkowski space. The Unruh
effect is a direct
consequence of the fact that the particle content of a field
theory is very frame dependent \cite{U,F}. Recently, Bell and
Leinaas \cite{BL} raised the very interesting possibility of
interpreting the observed depolarization of electrons in
particle accelerators in terms of the Unruh effect. They argued
that electrons could be used as sensitive thermometers because
of the fact that
the coupling between the spin and the magnetic field induces a
splitting between the ``spin up'' and ``spin down'' levels. Thus, the
observed depolarization of electron beams might be
interpreted in the electron's rest frame as due to the thermal
bath predicted by Unruh.
Since electrons in {\em linear accelerators} do not
have time to reach equilibrium in the polarization distribution,
Bell and Leinaas decided to consider electrons in {\em storage
rings}  \cite{BL}.
In this case, ultra-relativistic
electrons following uniform circular motion
experience in their rest frame an ``effective'' thermal bath
characterized by a temperature which differs from the original Unruh
temperature by a numerical factor, $\pi / \sqrt 3$ \cite{T,LP}.
Although other effects \cite{CPL} may play an important role in this
phenomenon, Bell and Leinaas' results are in good
agreement with experiment.  At LEP
conditions, electrons are typically accelerated at
$a=2.9 \times 10^{23} m/s^2$, which corresponds to an Unruh temperature
of $\hbar a/2\pi c k= 1200K$. This is only 4 times larger than typical
laboratory temperatures. In this vein, it would be
desirable to consider a detector accelerated in a background
thermal bath rather than in the inertial vacuum
in order to investigate in what extent
finite-temperature corrections should be taken into account when
testing the Unruh effect under real laboratory conditions.
The detector excitations will represent
the electron depolarizations, since both ones
share the common feature of being
two-level systems  \cite{BL}.
We show that in real accelerator conditions the major contribution
to the detector's response  comes from the inertial vacuum rendering the
contribution due to the presence of the background thermal bath
unimportant. It corroborates the usual assumption, when
testing the Unruh effect in storage rings, of considering the
electrons as being accelerated in the Minkowski vacuum \cite{BL,DIIM}.

The paper is organized as follows: In Section \ref{sec:inertial}
we study {\em inertial} detectors
evolving in a background thermal bath and show that due to time
dilatation the faster the detector moves, the less it interacts with the
thermal bath. In Section \ref{sec:uniform} we replace the
inertial detectors by {\em uniformly accelerated} ones, and calculate
finite-temperature corrections in the
detector's excitation rate due to the external thermal bath.
In Section \ref{sec:circ} we consider detectors moving
{\em circularly with constant speed,} and discuss our results in
connection with the proposal of testing the Unruh effect in storage
rings. Final conclusions are summarized in Section
\ref{sec:conclusions}. Natural units will be used ($\hbar = c = k =
1$) unless stated otherwise, and the signature adopted is $(+---)$.

\section{Inertial detectors in a background thermal bath}
\label{sec:inertial}
We will show in this section that the faster a detector moves in
a background thermal bath the less the detector interacts with the bath. This
is so because time dilatation induces a fast moving detector to
interact preferentially with low frequency modes.  Although a
thermal bath is rich of low frequency modes, the phase space
volume element ($ \propto \omega^2 d\omega$) suppresses infrared
contributions.

Let us begin considering an Unruh-DeWitt detector \cite{U,DW}.
It is basically a two-level device which may be either in the
ground state $\vert E_0\rangle$ or in the excited state $\vert
E\rangle$. The detector will be described by a monopole $\hat
m(\tau)$ coupled to a massless scalar field $\hat{\phi} (x^\mu)$
through the interaction action
\begin{equation}
S_I = \int_{-\infty}^{+\infty} d\tau\; c(\tau ) \hat{m} (\tau ) \hat{\phi}
\left[ x^\mu (\tau )\right] ,
\label{SI}
\end{equation}
where $x^\mu (\tau )$ is the detector's world line and $\tau $ is its proper
time. Here $c(\tau )$ is a switching function through which the
detector is turned on/off, and plays the role of a small
coupling parameter. In this section it will be enough to
consider a permanently switched on detector, {\em i.e.,} $c(\tau ) = c_0 =$
const. In the Heisenberg picture the monopole operator is time evolved as
\begin{equation}
\hat m(\tau ) = e^{i\hat{H}_0\tau } \hat{m} (0)e^{-i\hat{H}_0\tau } ,
\label{MTAU}
\end{equation}
where $\hat{H}_0 \vert E\rangle = E \vert E \rangle$ for any
detector's energy eigenstate $\vert E\rangle$ .

The amplitude for the detector to be excited and simultaneously
absorb a particle $\vert {\bf k} \rangle$ is
\begin{equation}
{\cal A}^{\rm exc}_{\rm abs} =
          \langle 0 \vert \otimes \langle E
          \vert S_I \vert
          E_0 \rangle \otimes \vert {\bf k} \rangle .
\label{AEX1}
\end{equation}
Using the expansion of the scalar field in plane waves (see,
{\em e.g.,} \cite{IZ}), and assuming that our detector follows
an inertial world line $x = y = 0$; $z = vt$; $t =
\tau/ \sqrt{1-v^2}$ (where $v=\vert {\bf v} \vert$ is the detector's
speed with respect to the background thermal bath), we obtain
\begin{equation}
{\cal A}^{\rm exc}_{\rm abs} =
          \frac{c_0}{\sqrt{4\pi\omega}}
          \delta \left[
          \Delta E - \frac{\omega - k_zv}{\sqrt{1-v^2}}
                 \right],
\label{AEX2}
\end{equation}
where $\Delta E = E - E_0$, and $\omega = \vert {\bf k} \vert$.
We will assume the selectivity
$\langle E \vert \hat{m}(0) \vert E_0 \rangle \equiv 1$
since it only depends on the internal details of the detector,
and it can be always factorized out.
The amplitude for the detector to be excited and simultaneously
emit a particle into the vacuum
vanishes due to energy conservation. Thus, {\em at
tree level,} the {\em total} excitation rate per {\em total}
proper time $T^{tot}$ of the detector will be
\begin{equation}
\frac{{\cal P}^{\rm exc}}{T^{tot}} =
                       \frac{1}{T^{\rm tot}}
                       \int d^3 {\bf k}
                       \vert {\cal A}^{\rm exc}_{\rm abs}\vert^2
                       \left[ \frac{1}{e^{\beta \omega } - 1} \right],
\label{PT}
\end{equation}
where $T^{tot} = 2 \pi \delta (0)$ \cite{IZ},
and we have added into brackets the usual absorption weight
associated with a thermal bath at a temperature $\beta^{-1}$.

As a consequence of (\ref{AEX2}), fast moving detectors will
only interact with low frequency modes. The very behavior of the
detector will be determined in (\ref{PT}) by the competition
between the thermal bath, which is rich of
low frequency modes, and the phase space volume element
$(\propto \omega^2d\omega )$ which tends to suppress infrared
contributions. Substituting
(\ref{AEX2}) in (\ref{PT}), and
performing the integrals, we obtain
\begin{equation}
\frac{{\cal P}^{\rm exc}}{T^{tot}} = \frac{c^2_0
\beta^{-1} \sqrt{1-v^2}}{4 \pi v} \ln\left[
         \frac{1-e^{-\beta \Delta E \sqrt{1+v}/\sqrt{1-v}}}
         {1-e^{-\beta \Delta E \sqrt{1-v}/\sqrt{1+v}}}
   \right].
\label{Final1}
\end{equation}
In the limit $v \to 0$ the detector responds with a Planckian
spectrum
$$
\frac{{\cal P}^{\rm exc}}{T^{\rm tot}}
= \frac{c_0^2}{2\pi }
\frac{\Delta E}{e^{\beta \Delta E}-1} ,
$$
as expected, while as $v \to 1$ the excitation rate per
total proper time vanishes (see Fig. \ref{fig:iner}). This suggests that in
testing the Unruh effect in ultra-relativistic regimes as in
particle accelerators \cite{BL,DIIM}, background thermal contributions
should be small. This issue will be investigated in detail next.

\section{Uniformly accelerated detectors in a background thermal bath}
\label{sec:uniform}
The total excitation probability for a uniformly accelerated
detector evolving in a background thermal bath
characterized by a temperature $\beta^{-1}$ can be
written as \cite{BD}
\begin{equation}
{\cal P}^{\rm exc} = \int_{-\infty}^{+\infty} d\tau c(\tau )
           \int_{-\infty}^{+\infty} d\tau' c(\tau ' )
           \; e^{-i\Delta E(\tau - \tau ')}
           G^+_{\beta} [x^\mu (\tau ), x^\mu (\tau ' )] ,
\label{P1}
\end{equation}
where
\begin{equation}
G^+_{\beta} [x^\mu (\tau ), x^\mu (\tau ' )] =
-\sum_{n=-\infty }^{+\infty }
          \frac{(4\pi^2)^{-1}}{
          (t - t'-i\beta n -i\varepsilon )^2 - \vert {\bf x} - {\bf x}'\vert^2
                              } ,
\label{Sum1}
\end{equation}
is the Wightman function,
$x^\mu (\tau )$ is the world line of the accelerated detector,
and $\tau$ is its proper time.
The world line of a detector moving in the $z$-axis
with constant proper acceleration $a$ is
\begin{equation}
t=\frac{1}{a} \sinh a\tau, \;\;\; z= \frac{1}{a} \cosh a\tau, \;\;\;
x=y=0 .
\label{WL}
\end{equation}
Substituting (\ref{WL}) in the Wightman function (\ref{Sum1})
we obtain
\begin{equation}
G^+_\beta (\tau, \tau') = -\frac{a^2}{16\pi^2}
\sum_{n=-\infty}^{+\infty}
\frac{1}{\left[\sinh a\Delta \tau /2 +i(n\beta a-\epsilon) e^{a\xi}/2\right]
       \left[\sinh a\Delta \tau /2 +i(n\beta a-\epsilon) e^{-a\xi}/2\right]} ,
\label{OI}
\end{equation}
where $\Delta \tau \equiv \tau - \tau '$, and $\xi \equiv (\tau + \tau ')/2$.
Using the following identity (the prime indicates that $n=0$ is
excluded from the sum)
\begin{equation}
{\sum_{n=-\infty}^{+\infty}}' \frac{1}{(A+iBn)(A+iCn)} =
\frac{2}{(C-B)B}\sum_{n=1}^{+\infty}\frac{1}{(A^2/B^2 + n^2)} \;,
\label{Sum2}
\end{equation}
in conjunction with \cite{GR}
\begin{equation}
\sum_{n=1}^{+\infty}\frac{1}{x^2 + n^2} = \frac{\pi}{2x}\coth{\pi x} -
\frac{1}{2x^2} ,
\label{Sum3}
\end{equation}
we can cast (\ref{OI}) in the form
\begin{equation}
G^+_\beta (\tau ,\tau ') = G_{\rm vac}^+(\Delta \tau )
                  + G_{\rm ther}^+(\Delta \tau ,\xi ) .
\label{G1}
\end{equation}
The first term in (\ref{G1}) corresponds to the pure vacuum
contribution \cite{BD}:
\begin{equation}
G_{\rm vac}^+(\Delta \tau ) =
            -\frac{a^2}{16\pi^2}
            \sinh^{-2} (a\Delta \tau/2 - i\varepsilon) ,
\label{GV}
\end{equation}
while the second term corresponds to the background thermal bath contribution:
\begin{eqnarray}
G_{\rm ther}^+(\Delta \tau ,\xi ) =& &
               \frac{a^2}{16\pi^2}\sinh^{-2} (a\Delta \tau/2)
              + \frac{a}{16\pi \beta \sinh{a\xi} \sinh(a\Delta \tau /2)}
              \nonumber \\
              & & \times
              \left[
              \coth\frac{2\pi \sinh (a\Delta \tau/2)}{a\beta e^{-a\xi}}
                               -
              \coth\frac{2\pi \sinh (a\Delta \tau/2)}{a\beta e^{a\xi}}
              \right] .
\label{GT}
\end{eqnarray}
The fact that $G_{\rm ther}^+$ depends on
$\xi$ reflects the fact that this is a non-stationary situation.
Notice that $G^+_{\rm ther} (\Delta \tau ,\xi )$ does not
diverge at any point. In particular $G^+_{\rm ther}(\Delta \tau
=0, \xi) =1/12\beta ^2$. Asymptotically
$G^+_{\rm ther}(\Delta \tau, \xi)$ behaves as
(see Fig. \ref{fig:accel})
\begin{equation}
G^+_{\rm ther} [\vert\Delta \tau \vert \gg (a^{-1},\beta);\;\; \xi]
                \sim e^{-a\vert \Delta \tau \vert}, \;\;\;
G^+_{\rm ther} [\Delta \tau \neq 0 ;\;\; \vert \xi \vert \gg (a^{-1},\beta)]
                \sim e^{-a\vert \xi \vert}.
\label{Gbig}
\end{equation}
Clearly, $G^+_{\rm ther}$ vanishes in the limit
{\mbox {$\beta \to +\infty$}}, and thus $G^+_{\beta \to +\infty}
(\tau, \tau') = G^+_{\rm vac} (\Delta \tau)$.

Now, we are ready to investigate the total excitation rate,
${\cal P}^{\rm exc} = {\cal P}^{\rm exc}_{\rm vac}
                      + {\cal P}^{\rm exc}_{\rm ther}$,
of a detector uniformly accelerated in a background thermal
bath. In order to be
realistic we shall consider the detector as being switched on
only during a finite period of proper time $\vert \tau \vert <
T_0/2 $, where $T_0 = {\rm const} \in {\bf R}_+$.
Concerning the pure vacuum contribution, ${\cal P}^{\rm exc}_{\rm vac}$,
it could be calculated
for some continuous $c(\tau )$ by
letting (\ref{GV}) into (\ref{P1}). Notwithstanding, we will use
directly the results of Ref. \cite{HMP} where the calculations
were performed in the detector's rest frame.
The excitation probability for a detector uniformly accelerated
in the {\em inertial} vacuum, and kept switched on for long enough,
$T_0 \gg a^{-1}, \Delta E^{-1}$, is
\begin{equation}
{\cal P}^{\rm exc}_{\rm vac} \approx \frac{c_0^2}{2\pi }
\frac{\Delta E}{e^{2\pi \Delta E/a}-1} T_0 .
\label{PV}
\end{equation}
Notice the linear dependence with $T_0$
(see Fig. \ref{fig:response}), which is exactly what one should expect
due to the Unruh effect \cite{BD}. Here $c_0 = {\rm const}$ is the coupling
constant
between the field
and the monopole while the detector is kept switched on.
In the regime considered above, the detailed form of $c(\tau )$
is not important. The only restriction is that $c(\tau ) \in
C^0$,
since discontinuities in $c(\tau )$ would result in ultraviolet
divergences \cite{HMP}.

The thermal correction, ${\cal P}^{\rm exc}_{\rm ther}$,
on the pure vacuum term (\ref{PV})
will be obtained by introducing (\ref{GT}) in (\ref{P1})
\begin{equation}
{\cal P}^{\rm exc}_{\rm ther} = c_0^2 \int_{-T_0/2}^{+T_0/2} d\tau
           \int_{-T_0/2}^{+T_0/2} d\tau'
           \; e^{-i\Delta E \Delta \tau }
           G^+_{\rm ther} (\Delta \tau, \xi ) ,
\label{P1T}
\end{equation}
where we have already considered the fact that the
detector is kept switched on only during a finite amount of proper time
$T_0$. The integrals above were solved numerically for $\Delta
E= 9.7 \times 10^{14} s^{-1}$, $ a = 9.7 \times 10^{14} s^{-1}$,
$\beta = 2.5 \times 10^{-14} s$, and plotted as a function of
$T_0$ in Fig. \ref{fig:response}. These values of $\Delta E$, $a$, and $\beta$
have a clear physical motivation.
Electrons in particle accelerators have their spin coupled to
the magnetic field. It induces a splitting of the ``spin up'',
and ``spin down'' levels. The energy gap associated with such a
splitting is $\Delta E=2 \vert \mu \vert \vert {\bf B} \vert$,
where $\mu \approx e/2m_e$ is the electron's magnetic moment
(it is assumed the gyromagnetic factor to be $g=2$), and
${\bf B}$ is the magnetic field.
Following \cite{BL} we consider the depolarization of an
accelerated electron as representing the excitation of the
detector, since both ones share the common feature of being
two--level systems. More detailed
calculations are not supposed to change the order of magnitude of the
results obtained, and consequently our conclusions.
At LEP conditions  an
electron has a typical Lorentz factor of $\gamma \equiv
(1-v^2)^{-1/2}= 10^5$, and proper acceleration of $a = 2.9 \times
10^{23}m/s^2$, which corresponds to $a = 9.7 \times 10^{14}s^{-1}$ in
natural units.  The energy gap between the two spin levels is
$\Delta E \approx a = 9.7 \times 10^{14} s^{-1}$ \cite{BL}.
The lab-time
for building up the polarization is about 2 hours, which
corresponds to $ 7.2 \times 10^{-2} s$ in the electron's
proper time. Finally, the background
thermal bath has a temperature corresponding to $\beta = (300 K)^{-1}
\approx 2.5 \times 10^{-14} s$.
{}From Fig. \ref{fig:response} it is clear that the pure vacuum contribution,
${\cal P}^{\rm exc}_{\rm vac}$, increases with $T_0$ much
faster than the background thermal contribution, ${\cal P}^{\rm
exc}_{\rm ther}$.  This is a
consequence of the fact that $G^+_{\rm ther}$
decreases exponentially for large $\vert T_0 \vert$
(Fig. \ref{fig:accel}), except along
the  $\xi$-axis. Eventually, Fig. \ref{fig:response} just reflects
the fact that the faster the detector moves the less it interacts
with the background thermal bath as discussed
in Section \ref{sec:inertial}.

Corrections on ${\cal P}^{\rm exc}_{\rm vac}$
due to the background thermal contribution, ${\cal P}^{\rm
exc}_{\rm ther}$,  are only important for
particles accelerated during short periods of time. It might be
the case in some regimes of
heavy-ion collisions. Barshay and Troost
suggested that the large transverse acceleration $a$ of
projectile and target which occurs in high-energy hadronic collisions
is connected with the thermal emission of particles at Unruh
temperature of about $100 MeV \approx 10^{12} K$ \cite{BT}. In acceleration
regimes in which the Unruh temperature is smaller, background thermal
contributions may be important. In this vein it might be useful
to expand  $G^{+}_{\rm ther} (\Delta \tau, \xi )$
in terms of $\beta$ factors. For this purpose,
we have used  Eq.(1.411.8) of \cite{GR} in (\ref{GT}) obtaining
\begin{equation}
G_{\rm ther}^+(\Delta \tau ,\xi ) = \frac{1}{12\beta^2}
                                  -
                                    \sum_{n=2}^{+\infty}
                                    \frac{(4\pi )^{2n-2} B_{2n}
                                          \sinh^{2n-2}(a\Delta \tau /2)
                                          \sinh[(2n-1)a\xi ]}{
                                          (2n)! {a}^{2n-2} \beta^{2n}
                                           \sinh{a\xi}} ,
\label{GTE}
\end{equation}
where $B_n$ are the Bernoulli numbers, and
$\beta > 2 a^{-1}\vert e^{a\xi}
         \sinh (a\Delta \tau/2)
         \vert $.
In the next section we will analyze a detector moving
circularly, and discuss our results in connection with
the proposal of testing the Unruh effect in storage rings by
looking at the polarization distribution of the accelerated electrons.

\section{Circularly moving detectors in a background thermal bath}
\label{sec:circ}

The world line of a detector describing a circular motion with
radius $R$, and constant speed $v$ is
\begin{equation}
t=\gamma \tau \; ,\;\;
x=R \cos \omega \gamma \tau \; ,\;\;
y=R \sin \omega \gamma \tau \; ,\;\;
z=0,
\label{CEM}
\end{equation}
where $\omega = v/R$.
The proper acceleration of the detector is $a\equiv \sqrt{a_\mu
a^\mu} = v^2 \gamma^2 /R$, where $a^\mu = u^\nu \nabla_\nu u^\mu$.

Substituting (\ref{CEM}) in ({\ref{Sum1}), and using (\ref{Sum2})
and (\ref{Sum3}) we decompose the relevant Green function again in
a pure vacuum part, and in a background thermal part as
\begin{equation}
D^+_\beta (\Delta \tau) = D_{\rm vac}^+(\Delta \tau )
                  + D_{\rm ther}^+(\Delta \tau) ,
\label{CD1}
\end{equation}
where
\begin{equation}
D_{\rm vac}^+(\Delta \tau ) = (4\pi^2)^{-1}
\left[-\gamma^2 (\Delta \tau -i\epsilon)^2 + 4 v^4 \gamma^4
\sin^2 (a\Delta \tau /2v\gamma)/a^2\right]^{-1}  ,
\label{CDVAC}
\end{equation}
and
\begin{eqnarray}
D_{\rm ther}^+(\Delta \tau) &=& - \frac{1}{4\pi^2\gamma^2 \Delta \tau^2}
\left\{ \left[\frac{4v^4 \sin^2 (A \Delta \tau /2 v)}
                   {A^2 \Delta \tau^2} -1 \right]^{-1}
                   + \frac{\pi A \Theta \Delta \tau^2}
                                   {4 v^2 \sin(A\Delta \tau /2 v)}
                                   \right.
                   \nonumber\\
                   &\times& \left.
                       \left[ \coth\left(\pi\Theta \Delta \tau
                              \left(1-\frac{2 v^2}{A \Delta \tau}
                              \sin \frac{A\Delta \tau}{2v}
                              \right) \right)
                              -(v \to -v)
                       \right]
                       \right\} .
\label{CDTHER}
\end{eqnarray}
Here $A\equiv a/\gamma$, and $\Theta\equiv \gamma/\beta$ play
the role of an effective proper acceleration, and an effective
background temperature respectively. $D_{\rm ther}^+(\Delta
\tau)$ is finite everywhere. In particular, $\lim_{\Delta \tau
\to 0} D_{\rm ther}^+(\Delta \tau) = 1/12\beta^2$.
Asymptotically, $D_{\rm ther}^+(\Delta \tau >> 1) \sim (4\pi^2
\gamma^2 \Delta \tau^2)^{-1}.$ The fact that $D_{\rm \beta}^+$
does not depend on $\xi$ reflects the fact that this situation
is stationary.

In order to calculate the average vacuum excitation rate,
$d{\cal P}^{\rm exc}_{\rm vac} /dT$, for ultra-relativistic
detectors it is convenient to express (\ref{CDVAC}) as
\begin{equation}
D^+_{\rm vac} (\Delta \tau) = (4\pi^2 )^{-1} [-(\Delta \tau
                            -i\epsilon)^2 -a^2 \Delta \tau^4/12
                            + {\cal O}(\gamma^{-1})] .
\label{GVAC3}
\end{equation}
Substituting (\ref{GVAC3}) in (\ref{P1}) we obtain for $\gamma
>>1$ \cite{T}
\begin{equation}
\frac{d{\cal P}^{\rm exc}_{\rm vac}}{dT}
\approx \frac{c_0^2 a e^{-\sqrt {12} \Delta E/ a}}{{4\pi
\sqrt{12}}} .
\label{TAK}
\end{equation}
Thus at LEP we expect
$d{\cal P}^{\rm exc}_{\rm vac}/dT
\approx 7.0 \times 10^{11} c_0^2 $.

In order to compare this result with the average  thermal
contribution, $d{\cal P}^{\rm exc}_{\rm ther} /dT$,
we substitute  (\ref{CDTHER}) in (\ref{P1}) obtaining
\begin{equation}
\frac{d{\cal P}^{\rm exc}_{\rm ther}}{dT} = c_0^2
           \int_{-\infty}^{+\infty} d(\Delta\tau)
           \; e^{-i\Delta E \Delta \tau}
           D^+_{\rm ther} (\Delta \tau)  .
\label{FIN}
\end{equation}
Evaluating numerically this expression with LEP values we obtain
$d{\cal P}^{\rm exc}_{\rm ther}/dT \approx 3 \times 10^{8} c_0^2 $.
Thus, after the equilibrium in
the polarization distribution is reached the pure vacuum
contribution is expected to be about 3 orders of magnitude
larger than the background thermal contribution. Before
concluding, we notice that for {\em ``quasi-inertial''}
detectors, i.e. $A << \Theta, \Delta E$, (\ref{FIN}) can be
approximated by the inertial formula  (\ref{Final1}). In
particular, in the limit $A \to 0$, $d{\cal P}^{\rm exc}_{\rm ther}/dT$
turns out to be exactly (\ref{Final1}). At LEP
conditions we have $A=9.7 \times 10^9 s^{-1}$,
$\Theta=4.0 \times 10^{18} s^{-1}$, and
$\Delta E = 9.7 \times 10^{14} s^{-1}$. Thus one could have used directly
(\ref{Final1}) to estimate the average thermal contribution
obtaining $d{\cal P}^{\rm exc}_{\rm ther}/dT
\approx 2.9 \times 10^{8} c_0^2 $.

\section{Conclusion}
\label{sec:conclusions}
We have derived the response of inertial and accelerated detectors
in a background thermal bath.
The faster the detector moves, the less important
will be the background thermal bath.
This is so because time dilatation induces the detector
to interact only with the low frequency modes present in the bath.
Although the thermal bath is rich of low frequency modes,
the phase space volume element suppresses infrared
contributions in the excitation probability.
Bell and Leinaas suggested that the depolarization of electrons
in storage rings could be explained through the Unruh effect,
i.e. due to the
appearance of a thermal bath in the electron's rest frame of
about $1200 K$. In their
analysis it is assumed that the electrons are accelerated in
the {\em inertial vacuum}. We have estimated whether
considering the fact that the electrons are actually
accelerated in a {\em background thermal bath} of about 300 K
would add or not any
substantial contribution in the depolarization rate. We obtain under LEP
conditions the interesting result that although the Unruh
thermal bath is only about 4 times hotter than the background
thermal bath, the term ${\cal P}^{\rm exc}_{\rm
ther}$  due to the external bath
is various orders of magnitude smaller than the pure vacuum
contribution ${\cal P}^{\rm exc}_{\rm vac}$.
Concerning the proposal of testing the Unruh effect in
storage rings, it corroborates the usual assumption of considering the
electrons as being accelerated in the inertial vacuum \cite{BL,DIIM}.
Notwithstanding, according to our results, we expect that
background thermal corrections will be relevant in
situations where non ultra-relativistic particles are
accelerated moderately for short periods of time.

\acknowledgments
We would like to acknowledge discussions with Dr. F. Kokubun in
the early stages of this work, and our gratitude to Dr. A.
Higuchi for his suggestions in the final manuscript.
This article was supported by Coordenadoria de
Aperfei\c coamento de Pessoal de N\'\i vel Superior (S.C.), and by Conselho
Nacional de Desenvolvimento Cient\'{\i}fico e Tecnol\'ogico (G.M.).

\begin{figure}
\protect
\caption{The excitation probability of an inertial detector
moving with speed $v$ in a thermal bath characterized by a
temperature {\protect $\beta^{-1}$} is plotted. The faster the detector moves
the less it
interacts with the thermal bath. Here  we have used
$\beta = 2.5 \times 10^{-14} s$,
and $\Delta E = 9.7 \times 10^{14} s^{-1}.$}
\label{fig:iner}
\end{figure}

\begin{figure}
\protect
\caption{$G^+_{\rm ther}
(\Delta \tau, \xi)$ is finite everywhere, and asymptotically
it decreases exponentially
except along the $\xi$-axis on which
it is constant. Notice that \protect $G^+_{\rm ther}
(\Delta \tau, \xi)$ is completely symmetric in the other quadrants.
Here we have used  $\beta = 2.5 \times 10^{-14} s$, and $a =
9.7 \times 10^{14} s^{-1}.$ }
\label{fig:accel}
\end{figure}

\begin{figure}
\protect
\caption{${\cal P}^{\rm exc}_{\rm vac}$ and ${\cal P}^{\rm
exc}_{\rm ther}$ are ploted as a function of $T_0$.
${\cal P}^{\rm exc}_{\rm vac}$, which is represented with a
full line, increases linearly for large $T_0$,
while ${\cal P}^{\rm exc}_{\rm ther}$, which is
represented with a dotted line, increases much
slower. This is a consequence of the fact that $G^+_{\rm ther}$
decreases exponentially for large $T_0$, except along the
``monorail'' $\Delta \tau =0$. Eventually, it reflects
the fact that the faster
the detector moves with respect to the background thermal bath,
the less the detector interacts with it. Here we have used
$\beta = 2.5 \times 10^{-14} s$, $a = 9.7 \times 10^{14} s^{-1},$ and
$\Delta E = 9.7 \times 10^{14} s^{-1}.$}
\label{fig:response}
\end{figure}

\end{document}